# Content Addressable Memories and Transformable Logic Circuits Based on Ferroelectric Reconfigurable Transistors for In-Memory Computing

Zijing Zhao, Junzhe Kang, Ashwin Tunga, Hojoon Ryu, Ankit Shukla, Shaloo Rakheja, and Wenjuan Zhu

*Department of Electrical and Computer Engineering, University of Illinois at Urbana-Champaign, Urbana, IL 61801, USA*

**Abstract:**

As a promising alternative to the Von Neumann architecture, in-memory computing holds the promise of delivering high computing capacity while consuming low power. Content addressable memory (CAM) can implement pattern matching and distance measurement in memory with massive parallelism, making them highly desirable for data-intensive applications. In this paper, we propose and demonstrate a novel 1-transistor-per-bit CAM based on the ferroelectric reconfigurable transistor. By exploiting the switchable polarity of the ferroelectric reconfigurable transistor, XOR/XNOR-like matching operation in CAM can be realized in a single transistor. By eliminating the need for the complementary circuit, these non-volatile CAMs based on reconfigurable transistors can offer a significant improvement in area and energy efficiency compared to conventional CAMs. NAND- and NOR-arrays of CAMs are also demonstrated, which enable multi-bit matching in a single reading operation. In addition, the NOR array of CAM cells effectively measures the Hamming distance between the input query and stored entries. Furthermore, utilizing the switchable polarity of these ferroelectric Schottky barrier transistors, we demonstrate reconfigurable logic gates with NAND/NOR dual functions, whose input-output mapping can be transformed in real-time without changing the layout. These reconfigurable circuits will serve as important building blocks for high-density data-stream processors and reconfigurable Application-Specific Integrated Circuits (r-ASICs). The CAMs and transformable logic gates based on ferroelectric reconfigurable transistors will have broad applications in data-intensive applications from image processing to machine learning and artificial intelligence.

**Introduction:**

The system-level performance of modern computing platforms is dominated by the energy and latency cost of data communication between processors and memory. This challenge, known as the "memory bottleneck", is exacerbated for memory-intensive applications from the machine learning and artificial intelligence domains. In-memory computing has emerged in recent years to address this problem.[1-6] By processing the data directly within the memory, in-memory computing can provide higher energy efficiency and faster processing speed.[1-6] Content addressable memory (CAM) is one of the promising in-memory computing circuits widely used in high-speed search applications.[2-3, 7-11] CAM can compare an input vector against a list of stored vectors in parallel and return the address of the matching entry or the distance between the input vector and the stored vector. In the conventional approach, the stored memory vectors need to be transferred to a computing unit to calculate the distances for a given query. For a single search operation in memory with *M* entries (each memory entry has *N* dimensions), the total number of required operations includes data transfer for *M* entries and *M*N* comparison computations, which would be very expensive, especially for large memory arrays. In contrast, CAM can perform the comparison directly within the memory cell and the comparison can be carried out in all the memory cells simultaneously. This enables the search results to be obtained in a single clock cycle, resulting in much greater energy efficiency and faster operation compared to traditional address-based memory. Due to the highly parallel search capabilities, CAM exhibits high potential for data-intensive applications, ranging from machine learning to routers and caches in processors.[2-3, 7-11] However, traditional SRAM-based CAM needs ten transistors (10T) per bit, which imposes severe limits on the area and energy efficiency.[12] Emerging non-volatile memories can simplify the CAM cell with fewer components. CAM cells with four transistors per bit (4T), two transistors and two memristors per bit (2T2R), or two ferroelectric transistors per bit (2T) have been demonstrated.[13-17] However, these implementations all rely on twin complementary circuits to represent the XOR/XNOR-like matching operation of CAM, which leads to extra hardware costs and energy consumption. In this paper, we propose a new type of CAM cell based on the ferroelectric reconfigurable transistor, which only needs one transistor per bit. By eliminating the complementary branches for the XOR/XNOR-like logic, the circuit density and energy efficiency can be significantly enhanced.

Data-stream processor is another interesting approach to address the memory bottleneck problem. In data-stream processors, the configurable logic blocks are connected as a pipeline, and large datasets are passed between logic blocks without intermediate stops at memory units.[18-19] By reducing data movement to memories, data-steam processors show superior processing speed, handling millions of tasks per clock cycle.[19] Traditional data-stream processors are mainly based on field-programmable gate arrays (FPGAs).[20] However, due to their large networks of wires and switch boxes, FPGAs exhibit much lower circuit density and higher power consumption compared to Application-Specific Integrated Circuit (ASIC).[21] In this paper, we show that reconfigurable logic gates can be implemented using ferroelectric reconfigurable transistors. Unlike FPGAs, in which the functionality of logic devices is fixed and reconfiguration of the circuit is realized by re-wiring, in these ferroelectric logic gates, the logic devices themselves can transform their behavior, therefore providing a simple yet powerful way to reconfigure circuit functionality. By eliminating the need for switch boxes as in FPGAs, circuits based on transformable devices will have higher integration density. Additionally, these ferroelectric reconfigurable logic circuits will be more energy efficient since the configuration is non-volatile.

In both of these circuits (CAM and transformable logic gates), reconfigurable transistors serve as the essential elements. Unlike the traditional CMOS transistor whose polarity (n-type or p-type) is fixed by the source/drain doping type, the polarity of the reconfigurable transistors can be dynamically changed even after fabrication. In the past, reconfigurable transistors based on electrostatic gating have been demonstrated using silicon nanowires and 2D semiconductors as channels.[22-28] The ultra-thin bodies of the 2D materials and silicon nanowires allow the electron or hole doping in the contact region to be effectively modulated by electrostatic gating, which enables the polarity switching in the transistor.[24-25, 29-30] However, these devices rely on a constant supply of external voltage to provide electrostatic doping, which leads to high energy consumption. Non-volatile reconfigurable transistors based on floating gates have been demonstrated, where the transistor polarity is controlled by the trapped charges on the floating gate.[1, 31] However, since the charges are injected to (or withdrawn from) the floating gate by tunneling through the thin gate dielectrics, frequency program/erase operations can lead to dielectric degradation or breakdown, which limits the endurance of the device. In addition, the current-based program and erase operation also impose significant power consumption. In contrast, ferroelectric-based non-volatile transistors are field-driven, which are more energy efficient and more durable. Recently, we have demonstrated non-volatile reconfigurable transistors based on ferroelectric materials, which show switchable polarity by local ferroelectric polarization.[32]

In this paper, we demonstrate 1-transistor-per-bit CAMs and transformable logic gates based on ferroelectric reconfigurable transistors. By exploiting the switchable polarity of the ferroelectric Schottky barrier transistors, the search function in a 1-bit CAM is implemented using a single transistor. By eliminating the need for complementary circuits, these CAMs based on ferroelectric reconfigurable transistors are much more efficient in circuit density and energy consumption than traditional CAMs. In addition, 2-bit CAMs with NAND- and NOR-arrays are also developed. The NOR-array CAMs not only can perform the match operation but also can measure the Hamming distance between the input query and stored entries. Furthermore, logic invertors and NAND/NOR dual-function logic gates are also demonstrated based on ferroelectric reconfigurable transistors, which will have broad applications in data-stream processors and high-security hardware.

**Non-volatile reconfigurable transistors based on van der Waals materials**

The schematic of the non-volatile reconfigurable transistor is shown in Figure 1a. The design is based on a back-gated transistor, consisting of an embedded back gate, a layer of back gate dielectric, source and drain contacts, and a $MoTe_2$ channel. 2D semiconductor $MoTe_2$ has a small bandgap of 1.0 eV and similar electron and hole mobilities, which supports both n-type and p-type current transport.[33-35] The ferroelectric $CuInP_2S_6$ (CIPS) layer is placed on top of the $MoTe_2$ channel. Two programming gates on top of the source and drain contacts are used to dynamically tune the ferroelectric polarizations of the CIPS layer locally. Figure 1b shows an optical image of a typical device. The thicknesses of the $MoTe_2$ and CIPS flakes are 10.2 nm and 162 nm respectively as extracted from atomic force microscopy (AFM) measurement (Figure S1). A 30 nm $Al_2O_3$ layer is used as the back gate dielectric. Two transistors are fabricated on the same flake, ensuring consistent channel thickness between the two devices. The addition of programming gates forms a ferroelectric metal/CIPS/$MoTe_2$/metal capacitor, and the polarity of the ferroelectric layer is controlled by the voltage applied between the two metals. During the programming operation, a voltage pulse is applied to the programming gate, and the bottom metal electrodes (source and drain) are shorted to the ground. We note that the two programming gates of a single transistor can be wired together. Ferroelectric capacitors are often used in the gate stack of ferroelectric transistors, where the interfacial doping from ferroelectric polarization alters the threshold voltage of the transistor, such that the transistor becomes a memory cell that can store 1-bit information. The versatile non-volatile reconfigurable transistors can perform both memory and logic functions. The two ferroelectric capacitors on the source and drain stack allow the reconfigurable transistor to operate as memory. On the logic side, when both capacitors are programmed, the transistor can switch between unipolar n-type and p-type operations in contrast to the threshold voltage shift. Therefore, the polymorphic device can be used for both logic and memory applications.

When the ferroelectric CIPS is polarized, the interfacial doping induced in the adjacent $MoTe_2$ will alter $MoTe_2$'s Fermi level at the contact region. A negative voltage on the programming gate will introduce hole doping. The band diagram in Figure 1c illustrates that hole doping from CIPS effectively narrows the Schottky barrier width for hole injection. When the back gate introduces hole accumulation in the channel, the transistor is turned on and behaves as unipolar p-type. The band diagram for the electron-doped case is shown in Figure 1d. Similarly, when both source and drain are doped with electrons, the contacts facilitate electron conduction, and the transistor shows a unipolar n-type behavior. The typical transfer curves after positive pulses and negative pulses are shown in Figure 1e and Figure 1f, respectively. Note that both curves were measured after removing the program pulses, indicating that the configurations are non-volatile. The pulse height and width used are ±5V and 4s, respectively. N-type transfer curves are observed after positive pulses, confirming the additional doping is contributed by ferroelectric polarization. As the source-drain voltage increases, both electron and hole conductions are enhanced. The on- and off-state currents increase while the on-off ratio decreases. This is a characteristic behavior of the ambipolar transport in the Schottky barrier transistor, where the high drain bias aids the carriers with high energy to overcome the large Schottky barrier.[36-38] The output curves of the n-type and p-type transistors are shown in Figure 1g and 1h respectively. At high gate bias, the output curves of both n-type and p-type configurations show a linear dependence of drain current on drain voltage, indicating the output current is controlled by channel conduction. The Schottky barrier widths at the contacts are negligible under ferroelectric doping. The retention performance of the transistor is also measured and presented in Figure S2. For the transistor programmed in n-type, it can maintain a sizable on-off contrast for 5400s.

**1-transistor-per-bit CAMs based on ferroelectric reconfigurable transistors**

Utilizing the switchable polarity of reconfigurable transistors, we demonstrate CAM cells with only one transistor per bit. The operation of the CAM cell based on the ferroelectric reconfigurable transistor is illustrated in Fig. 2a. The back gate is connected to the search line and driven by input vectors, while the drain is connected to the match line for current sensing. The stored entry of the CAM cell is represented by the different non-volatile states of the ferroelectrics, which determines the polarity of the transistor. Note that reconfigurable transistors differ from traditional ferroelectric memory, as they use different polarities to encode information instead of relying on threshold voltage shifts. During the program operation, we write the logic "0" state into the CAM cell by setting the ferroelectric transistor to p-type and write the logic "1" state by setting the ferroelectric transistor to n-type. Figure 2b illustrates the transfer characteristics of the reconfigurable transistor at these two states. Next, we search logic "0" by applying a low search voltage to the back gate and search logic "1" by applying a high search voltage. With above write and search schemes, when the stored and search data match, the ferroelectric transistor will be turned on and the channel will have high conductance, as shown in the truth table of the CAM cell (Fig. 2c). For example, when the store state is "1" (n-type) and the search logic is "1" (high gate voltage), the transistor will be turned on. In this case, a discharge current will pull the match line down to a low voltage level. However, when search data does not match with the stored data, the ferroelectric transistor will be in the "Off" state and the channel conductance will be low. The measured results of the fabricated CAM cell based on reconfigurable transistor are shown in Fig. 2d. Here $V_{SL} = 4V$ represents searching "1" and $V_{SL} = -4V$ represents searching "0". Only in the case of matching between the search and stored data, can the drain current of the ferroelectric transistor be high, indicating that the CAM cell is functional. Here the XNOR function in the CAM cell is implemented using a single transistor, due to the switchable polarity of the ferroelectric reconfigurable transistor. By eliminating the need for the twin complementary circuit, the circuit density of the CAM cell based on the reconfigurable transistor will be significantly higher than traditional CAM based on non-volatile memory and SRAMs. In addition, since the configuration (stored data) is non-volatile, these CAM cells based on ferroelectric reconfigurable transistors are much more energy efficient than SRAM-based CAMs.

Furthermore, we demonstrate NAND and NOR CAM arrays based on non-volatile reconfigurable transistors. The naming of NAND and NOR refers to how we connect the match lines of multiple bits. In the NAND design, all NAND cells are connected in series as shown in Figure 3a. The search line voltage of $4V$ and $-4V$ are defined as logic input "1" and "0" respectively in NAND array. The match line is pre-charged to "High" initially during the data search. When all the CAM cells are matched, the match line is discharged to a low voltage level. On the other hand, the NOR array connects all the NOR cells in parallel as shown in Figure 3b. The search voltage of 4V and -4V are defined as logic input "0" and "1" respectively in NOR array, which is opposite to the NAND array (Fig. 3c). Under this scheme, any mismatched cell will lead to a discharge current that pulls the match line to a low voltage level, The conductance of the CAM array increases as the number of mismatched cells increases, making the match line discharge faster. When all cells are matched, the discharge current will be minimal, and the potential of the match line remains "High". We verified the functionality of the NAND and NOR CAM arrays by programming the two bits into different polarities and by searching the arrays (Figure 3d-f). The drive voltage on the shared match line is 2V for the NAND array and 1V for NOR array. Both the NAND and NOR arrays show distinctive current levels when the input voltage vector matches the programmed states. Figure 3c-e gives the output currents under different inputs for a 2-bit NAND array programmed in n-n, p-p, and n-p states. In the NAND design, the mismatched states are all in low conductance states, giving a clear contrast to the matched state. Figure 3g-i presents the output currents for NOR array under the same measurements. In the case of the NOR array, the currents show three distinctive groups in all three n-n, p-p, and n-p programming states. The channel conductance is highest when both cells are in a matched state and is lowest when both cells are in mismatched states. Here, the current level in the match line depends on how many bits are in the matched states. Therefore, the discrete current levels also give the Hamming distance between the query data and the stored vectors in an analog way. The NOR array can also support partial matching, which is demonstrated in Figure S3. Figure S3b and S3c show the current in the match line as a function of search line voltages in n-n and p-p programmed NOR array. Aside from the regular search line voltages 4V and -8V for searching logic "0" and "1", applying a third voltage at -2V shows low currents for both n- and p-type transistors, therefore represents matching for this bit regardless of the polarization of the transistor. This mechanism can be used to mask out certain bits in an array and perform a partial match.

**Reconfigurable logic blocks design**

The switching of n-type and p-type configurations in transistors can host versatile functionality in CMOS circuits. Invertors are a fundamental CMOS logic block widely used for inverting or buffering digital signals. The circuit diagram of an inverter is shown in Figure 4a. The two transistors are configured into n-type and p-type. The gate and drain terminals of the two transistors are connected as the input voltage and output voltage respectively. The supply voltage $V_{DD}$ and ground are connected to the transistors' source terminals. The transfer curves of the invertor under different supply voltages are shown in Figure 4c. The circuit gives correct logic output for all supply voltages, and a sharp transition is observed when the input voltage switches from low to high. The gain is extracted from the derivative of the voltage transfer curve, and the maximum gain is 1.2 at a $V_{DD}$ of 3V. The gain of the invertor is strongly affected by the on-off current ratio of n-type and p-type transistors. Therefore, scaling the device dielectric and the supply voltage can improve the gain of the inverter. At $V_{BG}$=-6V, the saturated output voltage is slightly lower than the supply voltage, which is limited by the hole current ratio between the n-type and p-type transistors. In CMOS circuits, the p-type and n-type transistors have different mobilities, so tuning the transistor's width is necessary to match the drive strength. In the case of reconfigurable transistors, the close on-state currents of n-type and p-type transistors imply that one can use the same size for both transistors, regardless of their polarity.

More complex CMOS gates consist of pull-up and pull-down networks, built from p-type and n-type transistors with dual topology. The pull-up network has the inverted output of the pull-down network. The pull-up network is transparent to logic 1 exactly when the pull-down network blocks logic 0. When switching all p-type transistors to n-type, the pull-up network becomes a pull-down network, and the new network with the old topology now represents the dual function of the original pull-up network. The switching of pull-up and pull-down networks can be used to implement reconfigurable logic gates. Figure 4b shows the diagram of a reconfigurable NAND/NOR logic gate. When the transistors $T_1$ and $T_2$ switch from n-type to p-type, and the supply voltage switches with the ground, the function of the logic gate changes from NAND to NOR. The NAND/NOR logic gate operation is demonstrated with non-volatile reconfigurable transistors. Figure 4c-d show the static output voltage measured under different combinations of input logic. Here -4V and +4V are used to represent the logic 0 and 1 and the supply voltage is 1V. When the two series transistors are in the n-type configuration and *C* terminal is connected to $V_{DD}$, the output voltage matches the NAND function. On the other hand, when the two transistors are in the p-type configuration and *C* terminal is connected to GND, the function of the circuit matches the NOR logic. The effect of the load resistor on the output voltage of the logic gate is shown in Figure S4. For transistors programmed into n-n states, the margin between the "High" and "Low" output voltages increases with increasing load resistance, however, both output voltages are reduced. The resistors used for n-type and p-type operations are 5 MΩ and 12 MΩ respectively. The NAND/NOR reconfigurable gate can also be implemented with two transistors in parallel configuration (Figure S5). The circuit diagram is shown in Figure S5a. Two parallel-connected transistors serve as the pull-up/pull-down network. The load resistor is 9MΩ for all configurations, and -8V and 2V are used to represent logic 0 and 1. The effect of the supply voltage on output voltages for NOR gate is shown in Figure S5b. The gate shows a good contrast between the two output voltage levels for different input logic combinations. The detailed voltage sweeps on the A and B terminals in Figure S5c and S5d confirm the NAND/NOR logic gate operations under p-p/n-n configurations. The versatile reconfigurable logic gates provide an essential building block for programmable hardware. Unlike traditional FPGAs, in which the functionality of logic devices is fixed and reconfiguration of the circuit is realized by re-wiring, in these ferroelectric reconfigurable logic gates, the logic devices themselves can transform their behavior, therefore providing a simple yet powerful way to reconfigure circuit functionality. By eliminating the need for switch boxes as in FPGAs, circuits based on transformable devices will have higher integration density. Additionally, these ferroelectric logic circuits will be more energy efficient since the configuration is non-volatile. These reconfigurable logic gates can be used to implement data-stream processors, in which data is transferred between logic blocks without intermediate stops at the memory, thus enabling high processing speed.

**Conclusion**

In conclusion, we demonstrate high-density CAMs and transformable logic gates based on ferroelectric reconfigurable transistors. The polarity of the reconfigurable transistors can be switched between n-type and p-type by modulating the local ferroelectric polarization. By exploiting the switchable polarity of the Schottky contacts and the ambipolar transport of the vdW semiconductor, the search operation of the CAM cell can be implemented using a single ferroelectric transistor. By eliminating the complementary circuits, the CAMs based on ferroelectric reconfigurable transistors will have much higher area and energy efficiency compared to traditional CAMs. NAND and NOR arrays of CAM cells are also demonstrated by combining the transistors' matching lines in series or parallel. The array allows multi-bit matching in a single read operation. We found that the output current in the CAM with NOR array is discretized depending on the number of bits matched, enabling a direct calculation of the Hamming distance between "input" and "stored" vectors. Furthermore, we demonstrate logic gates that can be dynamically reprogrammed between NOR and NAND functionalities by switching the polarity of the transistors. These reconfigurable logic gates will enable reconfigurable ASICs, which can perform new or updated tasks, support new protocols, and resolve design issues. Furthermore, the device-level configurability will enable the system to achieve higher density and energy efficiency than traditional FPGAs. These CAMs and transformable logic gates will have broad applications in various applications ranging from machine learning, network routing, and image processing to artificial intelligence.

**Acknowledgment**

The authors would like to thank the support from Semiconductor Research Corporation (SRC) under Grant SRC 2021-LM-3042.

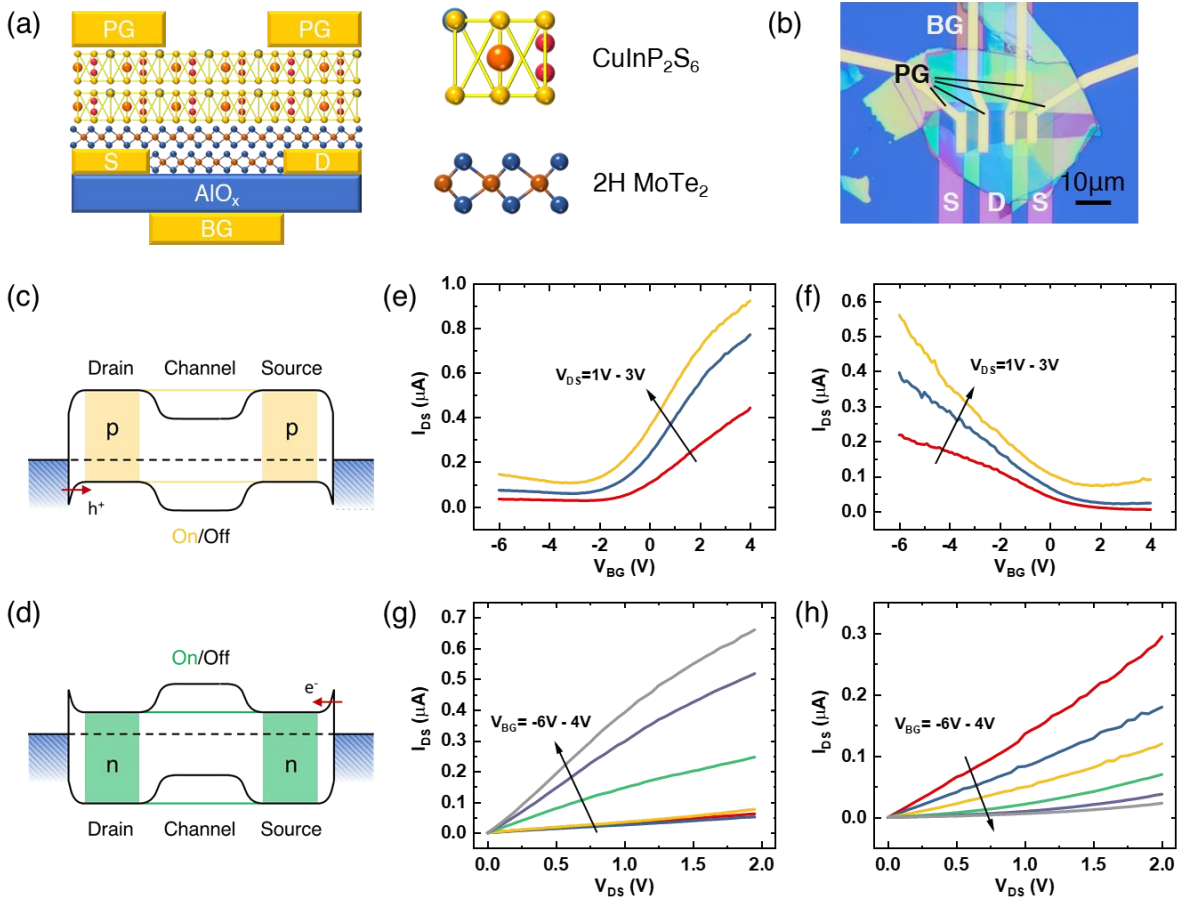

Figure 1. Ferroelectric reconfigurable transistor. (a) Schematic diagram of the non-volatile reconfigurable transistor. The source and drain electrodes consist of 3/15nm Ni/Au and the programming gates use a 20/60nm Ni/Au stack. The programming gate is extended by 0.6 μm to overlap with the channel. (b) Optical image of two transistors fabricated on the same MoTe$_2$ flake with a shared drain terminal. (c) (d) The channel band diagrams at equilibrium when additional hole (electron) doping is introduced in MoTe$_2$ by ferroelectric polarization. (e) (f) Transfer curves of a reconfigurable transistor in n-type and p-type modes measured at various drain voltages (from 1 V to 3 V with a 1 V interval). Unipolar behaviors are observed in both cases. (g) (h) Output curves of the reconfigurable transistor in n-type and p-type modes measured at various back gate voltages (from -6 V to 4 V with a 2 V interval). The output curves are linear at high gate doping indicating efficient electrostatic control from the gate.

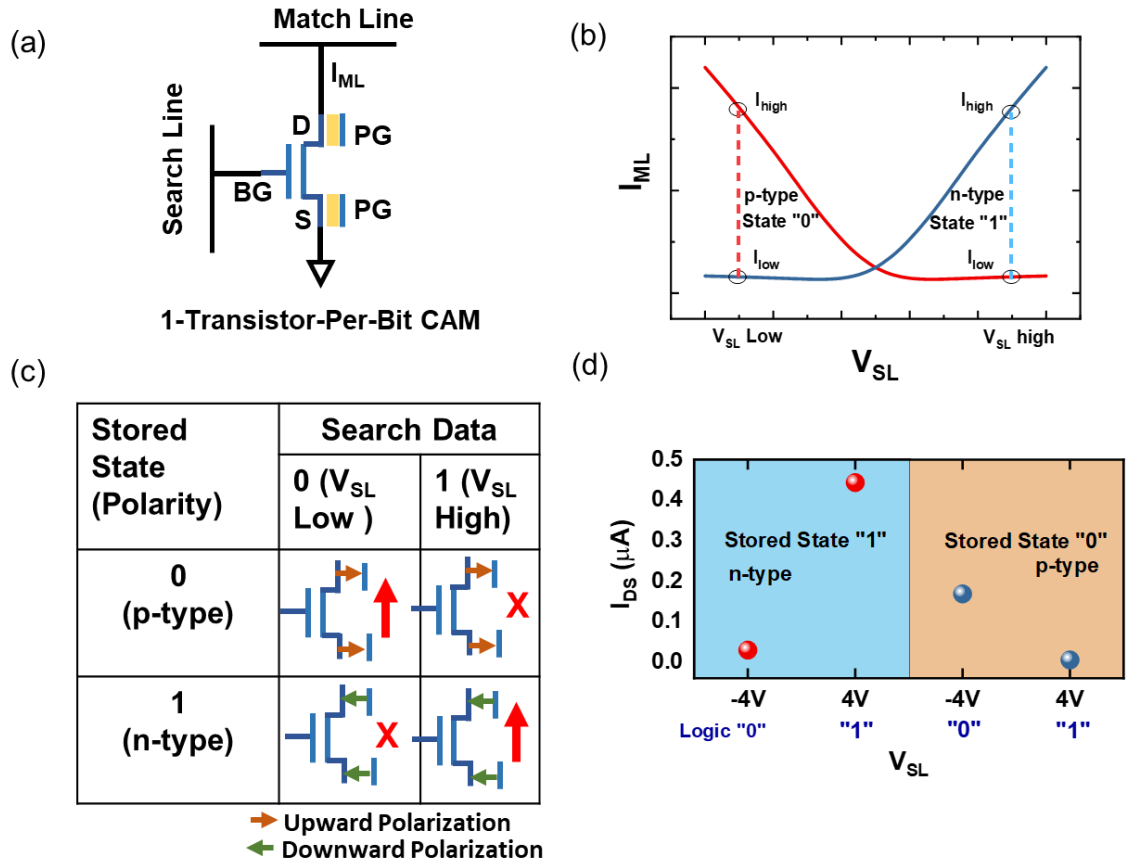

Figure 2. One-bit CAM based on the ferroelectric reconfigurable transistor. (a) Diagram of a ferroelectric reconfigurable transistor serving as a 1-bit CAM. The drain is connected to the matching line for sensing the output current, and the gate is connected to the search line for the input query. (b) Illustration of the drain current of the reconfigurable transistor with two different stored states under various search line voltages. (c) The truth table of the CAM cell based on the ferroelectric reconfigurable transistor. When the search data matches (mismatches) the stored state, the channel of the ferroelectric reconfigurable transistor will have high (low) conductance. (d) The drain current measured in a ferroelectric reconfigurable transistor under various combinations of the search input ($V_{SL}$) and the stored state, confirming that the output current can serve as an indicator for the matching and mismatching between the search data and stored state.

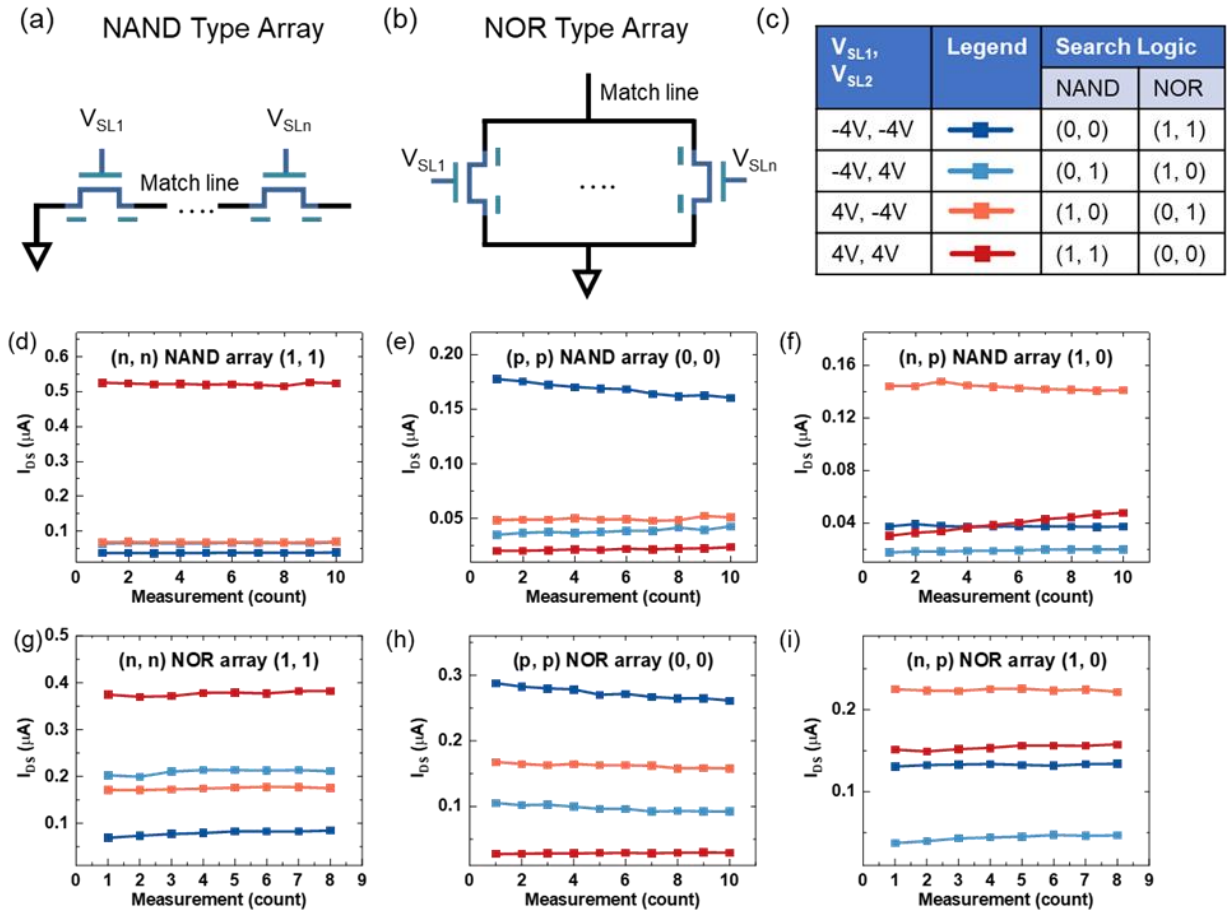

Figure 3. CAM arrays based on ferroelectric reconfigurable transistors. (a) The NAND array connects the transistors' match lines in series. The low conductance represents mismatch so that mismatching in any bit can be detected. (b) The NOR array connects the transistors' match lines in parallel. Here the high conductance represents a match. (c) Definition of search logic in NAND and NOR CAM arrays. (d-f) Measured drain current in a 2-bit CAM with NAND array. The 2-bit array is programmed in n-n (d), p-p (e), and n-p (f) states respectively. Currents are measured for all search line vectors. The highest current shows the matching states (1, 1), (0, 0), and (1, 0) in (d-f) respectively. (g-i) Measured drain current in a 2-bit CAM with NOR array. The 2-bit array is programmed in n-n (g), p-p (h), and n-p (i) states respectively. The highest current levels show the matching states (0, 0), (1, 1), and (0, 1) in (g-i) respectively. Partial matching also gives a distinctive current level, which can provide information on how many bits are in the matching states.

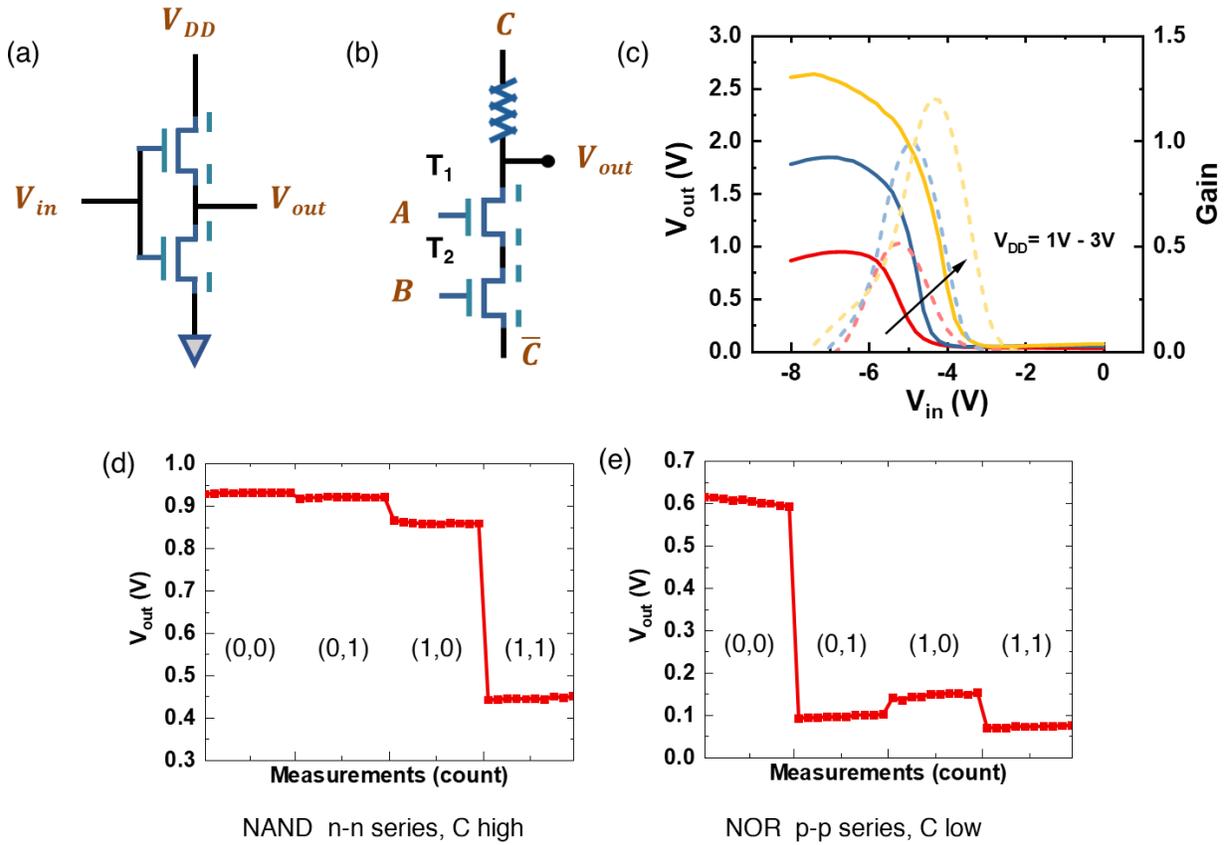

Figure 4. Logic gates based on ferroelectric reconfigurable transistors. (a) Circuit diagram of a CMOS invertor consisting of two reconfigurable logic transistors. The two transistors are configured into p-type and n-type respectively. (b) Illustration of reconfigurable NAND/NOR logic gates. (c) The voltage transfer curves and gain of an invertor with two reconfigurable transistors are measured with different supply voltages. When the transistors are configured into n-type and C is connected with the supply voltage, the output shows the NAND function, shown in (d). When the transistors are configured into p-type and C is connected to the ground, the output represents NOR logic, shown in (e).

**References:**

1. Marega, G. M.; Zhao, Y. F.; Avsar, A.; Wang, Z. Y.; Tripathi, M.; Radenovic, A.; Kis, A., Logic-in-memory based on an atomically thin semiconductor. *Nature* **2020**, *587* (7832), 72-+.
2. Yin, X.; Muller, F.; Laguna, A. F.; Li, C.; Ye, W.; Huang, Q.; Zhang, Q.; Shi, Z.; Lederer, M.; Laleni, N.; Deng, S.; Zhao, Z.; Niemier, M.; Hu, X. S.; Zhuo, C.; Kampfe, T.; Ni, K., Deep Random Forest with Ferroelectric Analog Content Addressable Memory. *arXiv* **2021**.
3. Li, H. Z.; Jin, H.; Zheng, L.; Liao, X. F., ReSQM: Accelerating Database Operations Using ReRAM-Based Content Addressable Memory. *Ieee T Comput Aid D* **2020**, *39* (11), 4030-4041.
4. Sebastian, A.; Le Gallo, M.; Khaddam-Aljameh, R.; Eleftheriou, E., Memory devices and applications for in-memory computing. *Nat Nanotechnol* **2020**, *15* (7), 529-544.
5. Chen, B.; Cai, F. X.; Zhou, J. T.; Ma, W.; Sheridan, P.; Lu, W. D., Efficient in-memory computing architecture based on crossbar arrays. *2015 Ieee International Electron Devices Meeting (Iedm)* **2015**.
6. Yin, L.; Cheng, R. Q.; Wen, Y.; Liu, C. S.; He, J., Emerging 2D Memory Devices for In-Memory Computing. *Adv Mater* **2021**, *33* (29).
7. Zhu, Q. L.; Graf, T.; Sumbul, H. E.; Pileggi, L.; Franchetti, F., Accelerating Sparse Matrix-Matrix Multiplication with 3D-Stacked Logic-in-Memory Hardware. *Ieee High Perf Extr* **2013**.
8. Kazemi, A.; Sharifi, M. M.; Laguna, A. F.; Muller, F.; Rajaei, R.; Olivo, R.; Kampfe, T.; Niemier, M.; Hu, X. S., In-Memory Nearest Neighbor Search with FeFET Multi-Bit Content-Addressable Memories. *Proceedings of the 2021 Design, Automation & Test in Europe Conference & Exhibition (Date 2021)* **2021**, 1084-1089.
9. Graves, C. E.; Li, C.; Sheng, X.; Miller, D.; Ignowski, J.; Kiyama, L.; Strachan, J. P., In-Memory Computing with Memristor Content Addressable Memories for Pattern Matching. *Adv Mater* **2020**, *32* (37).
10. Pedretti, G.; Graves, C. E.; Serebryakov, S.; Mao, R. B.; Sheng, X.; Foltin, M.; Li, C.; Strachan, J. P., Tree-based machine learning performed in-memory with memristive analog CAM. *Nat Commun* **2021**, *12* (1).
11. Li, C.; Müller, F.; Ali, T.; Olivo, R.; Imani, M.; Deng, S.; Zhuo, C.; Kämpfe, T.; Yin, X.; Ni, K. In *A scalable design of multi-bit ferroelectric content addressable memory for data-centric computing*, Int El Devices Meet, IEEE: 2020; pp 29.3. 1-29.3. 4.
12. Karam, R.; Puri, R.; Ghosh, S.; Bhunia, S., Emerging Trends in Design and Applications of Memory-Based Computing and Content-Addressable Memories. *P Ieee* **2015**, *103* (8), 1311-1330.
13. Xiong, X.; Kang, J. Y.; Liu, S. Y.; Tong, A. Y.; Fu, T. Y.; Li, X. F.; Huang, R.; Wu, Y. Q., Nonvolatile Logic and Ternary Content-Addressable Memory Based on Complementary Black Phosphorus and Rhenium Disulfide Transistors. *Adv Mater* **2022**, *34* (48).
14. Tong, L.; Peng, Z. R.; Lin, R. F.; Li, Z.; Wang, Y. L.; Huang, X. Y.; Xue, K. H.; Xu, H. Y.; Liu, F.; Xia, H.; Wang, P.; Xu, M. S.; Xiong, W.; Hu, W. D.; Xu, J. B.; Zhang, X. L.; Ye, L.; Miao, X. S., 2D materials-based homogeneous transistor-memory architecture for neuromorphic hardware. *Science* **2021**, *373* (6561), 1353-+.
15. Ni, K.; Yin, X. Z.; Laguna, A. F.; Joshi, S.; Dunkel, S.; Trentzsch, M.; Mueller, J.; Beyer, S.; Niemier, M.; Hu, X. S.; Datta, S., Ferroelectric ternary content-addressable memory for one-shot learning. *Nat Electron* **2019**, *2* (11), 521-529.
16. Yang, R.; Li, H. T.; Smithe, K. K. H.; Kim, T. R.; Okabe, K.; Pop, E.; Fan, J. A.; Wong, H. S. P., Ternary content-addressable memory with MoS2 transistors for massively parallel data search. *Nat Electron* **2019**, *2* (3), 108-114.
17. Li, C.; Graves, C. E.; Sheng, X.; Miller, D.; Foltin, M.; Pedretti, G.; Strachan, J. P., Analog content-addressable memories with memristors. *Nat Commun* **2020**, *11* (1).
18. Teubner, J.; Woods, L., *Data Processing on FPGAs*. 1st ed. 2013. ed.; Springer International Publishing: Cham, 2013.

19. Yoshida, H. Solving The Von Neumann Bottleneck With FPGAs. https://community.hitachivantara.com/blogs/hubert-yoshida/2019/02/01/solving-the-von-neumann-bottleneck-with-fpgas.
20. Schmit, H. H.; Cadambi, S.; Moe, M.; Goldstein, S. C., Pipeline Reconfigurable FPGAs. *Journal of VLSI signal processing systems for signal, image and video technology* **2000,** *24* (2), 129-146.
21. Kuon, I.; Rose, J., Measuring the Gap Between FPGAs and ASICs. *IEEE Transactions on Computer-Aided Design of Integrated Circuits and Systems* **2007,** *26* (2), 203-215.
22. Resta, G. V.; Balaji, Y.; Lin, D.; Radu, I. P.; Catthoor, F.; Gaillardon, P. E.; De Micheli, G., Doping-Free Complementary Logic Gates Enabled by Two-Dimensional Polarity-Controllable Transistors. *ACS Nano* **2018,** *12* (7), 7039-7047.
23. Wu, P.; Reis, D.; Hu, X. B. S.; Appenzeller, J., Two-dimensional transistors with reconfigurable polarities for secure circuits. *Nat Electron* **2021,** *4* (1), 45-53.
24. Larentis, S.; Fallahazad, B.; Movva, H. C. P.; Kim, K.; Rai, A.; Taniguchi, T.; Watanabe, K.; Banerjee, S. K.; Tutuc, E., Reconfigurable Complementary Monolayer MoTe2 Field-Effect Transistors for Integrated Circuits. *ACS Nano* **2017,** *11* (5), 4832-4839.
25. Nakaharai, S.; Yamamoto, M.; Ueno, K.; Lin, Y. F.; Li, S. L.; Tsukagoshi, K., Electrostatically Reversible Polarity of Ambipolar alpha-MoTe2 Transistors. *ACS Nano* **2015,** *9* (6), 5976-5983.
26. Heinzig, A.; Slesazeck, S.; Kreupl, F.; Mikolajick, T.; Weber, W. M., Reconfigurable silicon nanowire transistors. *Nano Lett* **2012,** *12* (1), 119-124.
27. Glassner, S.; Zeiner, C.; Periwal, P.; Baron, T.; Bertagnolli, E.; Lugstein, A., Multimode Silicon Nanowire Transistors. *Nano Lett* **2014,** *14* (11), 6699-6703.
28. Simon, M.; Liang, B.; Fischer, D.; Knaut, M.; Tahn, A.; Mikolajick, T.; Weber, W. M., Top-Down Fabricated Reconfigurable FET With Two Symmetric and High-Current On-States. *Ieee Electr Device L* **2020,** *41* (7), 1110-1113.
29. Chen, J.; Li, P.; Zhu, J. Q.; Wu, X. M.; Liu, R.; Wan, J.; Ren, T. L., Reconfigurable MoTe2 Field-Effect Transistors and Its Application in Compact CMOS Circuits. *Ieee T Electron Dev* **2021,** *68* (9), 4748-4753.
30. Pang, C. S.; Chen, C. Y.; Ameen, T.; Zhang, S. J.; Ilatikhameneh, H.; Rahman, R.; Klimeck, G.; Chen, Z. H., WSe2 Homojunction Devices: Electrostatically Configurable as Diodes, MOSFETs, and Tunnel FETs for Reconfigurable Computing. *Small* **2019,** *15* (41).
31. Sun, X. X.; Zhu, C. G.; Yi, J. L.; Xiang, L.; Ma, C.; Liu, H. W.; Zheng, B. Y.; Liu, Y.; You, W. X.; Zhang, W. J.; Liang, D. L.; Shuai, Q.; Zhu, X. L.; Duan, H. G.; Liao, L.; Liu, Y.; Li, D.; Pan, A. L., Reconfigurable logic-in-memory architectures based on a two-dimensional van der Waals heterostructure device. *Nat Electron* **2022,** *5* (11), 752-+.
32. Zhao, Z. J.; Rakheja, S.; Zhu, W. J., Nonvolatile Reconfigurable 2D Schottky Barrier Transistors. *Nano Lett* **2021,** *21* (21), 9318-9324.
33. Yin, L.; Zhan, X. Y.; Xu, K.; Wang, F.; Wang, Z. X.; Huang, Y.; Wang, Q. S.; Jiang, C.; He, J., Ultrahigh sensitive MoTe2 phototransistors driven by carrier tunneling. *Appl Phys Lett* **2016,** *108* (4).
34. Park, Y. J.; Katiyar, A. K.; Hoang, A. T.; Ahn, J. H., Controllable P- and N-Type Conversion of MoTe2 via Oxide Interfacial Layer for Logic Circuits. *Small* **2019,** *15* (28).
35. Chang, Y. M.; Yang, S. H.; Lin, C. Y.; Chen, C. H.; Lien, C. H.; Jian, W. B.; Ueno, K.; Suen, Y. W.; Tsukagoshi, K.; Lin, Y. F., Reversible and Precisely Controllable p/n-Type Doping of MoTe2 Transistors through Electrothermal Doping. *Adv Mater* **2018,** *30* (13).
36. Yang, L. M.; Charnas, A.; Qiu, G.; Lin, Y. M.; Lu, C. C.; Tsai, W.; Paduano, Q.; Snure, M.; Ye, P. D., How Important Is the Metal-Semiconductor Contact for Schottky Barrier Transistors: A Case Study on Few-Layer Black Phosphorus? *Acs Omega* **2017,** *2* (8), 4173-4179.
37. Das, S.; Demarteau, M.; Roelofs, A., Ambipolar Phosphorene Field Effect Transistor. *Acs Nano* **2014,** *8* (11), 11730-11738.
38. Das, S.; Appenzeller, J., WSe2 field effect transistors with enhanced ambipolar characteristics. *Appl Phys Lett* **2013,** *103* (10).